\documentclass[prl,superscriptaddress,twocolumn]{revtex4}

\usepackage{graphicx}
\usepackage{amsmath}

\begin{document}
\newtheorem{theorem}{Theorem}
\newtheorem{acknowledgement}[theorem]{Acknowledgement}
\newtheorem{algorithm}[theorem]{Algorithm}
\newtheorem{axiom}[theorem]{Axiom}
\newtheorem{claim}[theorem]{Claim}
\newtheorem{conclusion}[theorem]{Conclusion}
\newtheorem{condition}[theorem]{Condition}
\newtheorem{conjecture}[theorem]{Conjecture}
\newtheorem{corollary}[theorem]{Corollary}
\newtheorem{criterion}[theorem]{Criterion}
\newtheorem{definition}[theorem]{Definition}
\newtheorem{example}[theorem]{Example}
\newtheorem{exercise}[theorem]{Exercise}
\newtheorem{lemma}[theorem]{Lemma}
\newtheorem{notation}[theorem]{Notation}
\newtheorem{problem}[theorem]{Problem}
\newtheorem{proposition}[theorem]{Proposition}
\newtheorem{remark}[theorem]{Remark}
\newtheorem{solution}[theorem]{Solution}
\newtheorem{summary}[theorem]{Summary}    
\def\r{{\bf{r}}}
\def\i{{\bf{i}}}
\def\j{{\bf{j}}}
\def\m{{\bf{m}}}
\def\k{{\bf{k}}}
\def\kt{{\tilde{\k}}}
\def\mt{{\hat{t}}}
\def\mG{{\hat{G}}}
\def\mg{{\hat{g}}}
\def\mGa{{\hat{\Gamma}}}
\def\mS{{\hat{\Sigma}}}
\def\mT{{\hat{T}}}
\def\K{{\bf{K}}}
\def\P{{\bf{P}}}
\def\q{{\bf{q}}}
\def\Q{{\bf{Q}}}
\def\p{{\bf{p}}}
\def\x{{\bf{x}}}
\def\X{{\bf{X}}}
\def\Y{{\bf{Y}}}
\def\F{{\bf{F}}}
\def\G{{\bf{G}}}
\def\bG{{\bar{G}}}
\def\mbG{{\hat{\bar{G}}}}
\def\M{{\bf{M}}}
\def\V{\cal V}
\def\tchi{\tilde{\chi}}
\def\tx{\tilde{\bf{x}}}
\def\tk{\tilde{\bf{k}}}
\def\tK{\tilde{\bf{K}}}
\def\tq{\tilde{\bf{q}}}
\def\tQ{\tilde{\bf{Q}}}
\def\si{\sigma}
\def\ep{\epsilon}
\def\hep{{\hat{\epsilon}}}
\def\al{\alpha}
\def\be{\beta}
\def\ep{\epsilon}
\def\bep{\bar{\epsilon}_\K}
\def\mep{\hat{\epsilon}}
\def\up{\uparrow}
\def\de{\delta}
\def\De{\Delta}
\def\up{\uparrow}
\def\dwn{\downarrow}
\def\ksi{\xi}
\def\etha{\eta}
\def\product{\prod}
\def\goto{\rightarrow}
\def\switch{\leftrightarrow}

\title{Kinetic energy driven pairing}
\author{Th.A.~Maier}
\author{M.~Jarrell}
\affiliation{Department of Physics, University of Cincinnati, Cincinnati Ohio 45221, USA}
\author{A.~Macridin}
\affiliation{Department of Physics, University of Cincinnati, Cincinnati Ohio 45221, USA}
\affiliation{University of Groningen, Nijenborgh 4, 9747 AG Groningen, The Netherlands}
\author{C. Slezak}
\affiliation{Department of Physics, University of Cincinnati, Cincinnati Ohio 45221, USA}

\date{\today}

\begin{abstract}
Pairing occurs in conventional superconductors through a reduction of the 
electronic potential energy accompanied by an increase in kinetic energy, 
indicating that the transition is driven by a pairing potential. In the
underdoped cuprates, optical experiments show that pairing is driven by a 
reduction of the electronic kinetic energy. Using the Dynamical Cluster 
Approximation  we study the nature of superconductivity in a microscopic 
model of the cuprates, the two-dimensional Hubbard model. We find that 
pairing is indeed driven by the kinetic energy and that superconductivity 
evolves from an unconventional, spin-charge separated state, consistent 
with the RVB model of high-temperature superconductors.
\end{abstract}

\maketitle

The theory of superconductivity in the cuprates remains one of the most 
important outstanding problems in materials science.  Conventional 
superconductors are well described by the Bardeen-Cooper-Schrieffer (BCS) 
theory.  Here, the transition is due to the potential energy that electrons 
can reduce by forming Cooper pairs.  However, recent optical experiments show that 
the transition in the cuprates is due to a lowering of {\em{kinetic}} energy, 
suggesting that the mechanism for superconductivity in the cuprates is 
unconventional. 

In the BCS theory, pairing is a result of a Fermi surface instability that 
relies on the existence of quasiparticles in a Fermi-liquid.  The electrons
interact by exchanging phonons, the quanta of ionic vibrations of the crystal. 
Since this interaction leads to a net attractive force between 
electrons, the system can lower its potential energy by forming pairs
which have $s$-wave symmetry due to the local nature of the pairing interaction. 
These ``Cooper-pairs'' condense into a coherent macroscopic quantum state, 
insensitive to impurities and imperfections, and as a result, electricity 
can be conducted without resistance. 

The scattering of Cooper-pairs mediated by the attractive interaction leads 
to a reduction of its potential energy. To take advantage of this energy 
reduction, the electrons forming the pair have to occupy states outside the 
Fermi sea with an energy above the Fermi energy. As a result, pairing in 
conventional superconductors is always associated with an increase in 
kinetic energy which is overcompensated by the lowering of potential energy.

High-temperature cuprate superconductors (HTSC) are unconventional in various 
aspects and the pairing mechanism remains controversial.  The HTSC 
emerge from their antiferromagnetic parent compounds upon hole doping.  In 
the normal state of the weakly doped cuprates no quasiparticles are found, 
undermining the very foundation of BCS theory. It is widely believed that 
phonons cannot be responsible for pairing at temperatures as high as 
$160 {\rm K}$.  Consistently, the pairs have $d$-wave symmetry, instead of 
$s$-wave symmetry. Most significantly, new optical experiments 
\cite{marel:kineg,bontemps:transfer3} call for 
qualitatively different paradigms for HTSC. These experiments have shown 
that pairing in high-temperature superconductors  is driven by a reduction 
of the kinetic energy, not by an attractive potential as in the BCS theory.

Early in the history of HTSC it was realized that the two-dimensional (2D) 
Hubbard model in the intermediate coupling regime, where the Coulomb 
interaction between electrons is of the order of the bandwidth,  should 
capture the essential low-energy physics of the cuprates \cite{anderson:htsc}.  
However, these models lack exact solutions and approximative methods have 
to be applied.   
  
The foundation of the BCS theory relies upon a small parameter, the
ratio of the Debye-frequency to the Fermi energy $\omega_D/E_F$. One
of the complications of the purely electronic models of HTSC is the
lack of such a small parameter since the Coulomb repulsion between
electrons is roughly equal to their bandwidth. Perhaps the most
natural expansion parameter for these systems is the length scale of
antiferromagnetic spin correlations.  Neutron scattering experiments
confirm the presence of short-ranged antiferromagnetic correlations in
the doped cuprates up to length scales roughly equal to the mean
distance between holes, or roughly one lattice spacing in the
optimally doped cuprates with the highest transition temperature \cite{thurston:af}.  In
the dynamical cluster approximation
\cite{hettler:dca1,hettler:dca2,maier:dca1,jarrell:dca1} (DCA) we take
advantage of the short length-scale of antiferromagnetic correlations
and use it as a small parameter. The DCA reduces the complexity of the
problem by coarse-graining the $\k$-space on a scale $2\pi/L_c$.  As a
result, dynamical correlations up to a range $\sim L_c/2$ are treated
accurately while the physics on longer length scales is described on a
mean-field level.  The original lattice problem is mapped onto a
periodic cluster of size $N_c=L_c^D$ in $D$ dimensions embedded in a
host which has to be determined self-consistently. We solve the
cluster problem using quantum Monte Carlo and obtain dynamics from the
maximum entropy method \cite{jarrell:dca3}.

We present results of DCA calculations for the conventional 2D Hubbard 
model describing the dynamics of electrons on a square lattice. The model 
is characterized by a hopping integral $t$ between nearest neighbor sites  
and a Coulomb repulsion $U$ two electrons feel when residing on the same 
site. As the energy scale we set $t=0.25{\rm eV}$ so that the band-width 
$W=8t=2{\rm eV}$, and study the intermediate coupling regime $U=W$.  We 
study the dynamics on short length-scales by setting the cluster size to 
$N_c=4$, the smallest cluster size which allows for a superconducting 
phase with $d$-wave order parameter. This cluster size is large enough to 
capture the qualitative low-energy physics of the cuprate superconductors 
\cite{jarrell:dca2, maier:dca2}, while the solution retains some mean-field
behavior. 

\begin{figure}
\includegraphics*[width=90mm]{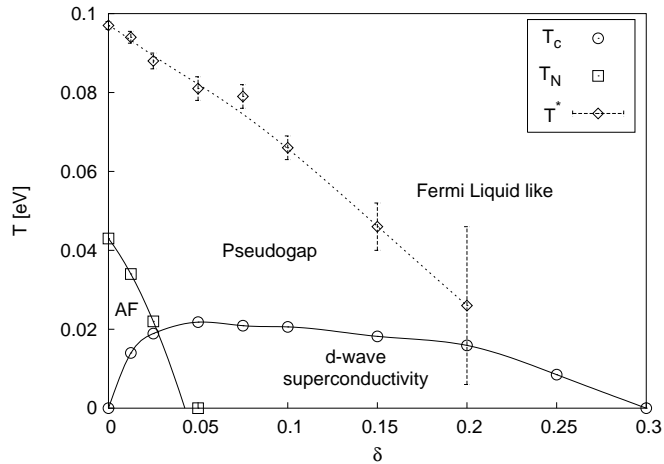}
\caption{DCA temperature-doping ($T$-$\delta$) phase diagram of the 2D 
Hubbard model when the Coulomb repulsion is equal to the bandwidth $U=W=2{\rm eV}$ 
for the DCA cluster size $N_c=L_c^2=4$. Hole doping renders the 
antiferromagnet  near zero doping ($\delta=0$) superconducting at 
low temperatures. In the normal state the electronic excitation spectrum 
shows a pseudogap below the crossover temperature $T^\star$. The errorbars on $T^\star$ indicate the difficulty at large doping in identifying the maximum in the spin-susceptibility which is used to determine $T^\star$. 
}
\label{fig:pd}
\end{figure}

\begin{figure}
\includegraphics*[width=45mm]{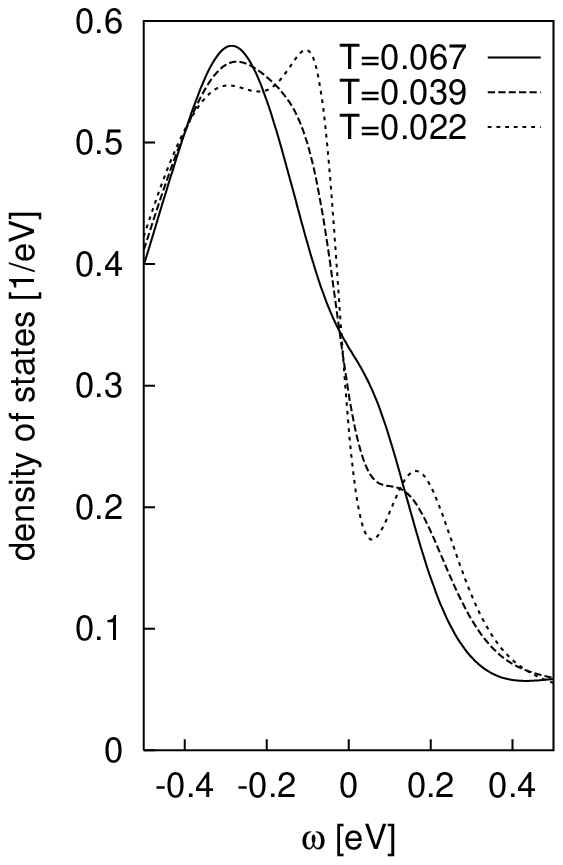}\includegraphics*[width=45mm]{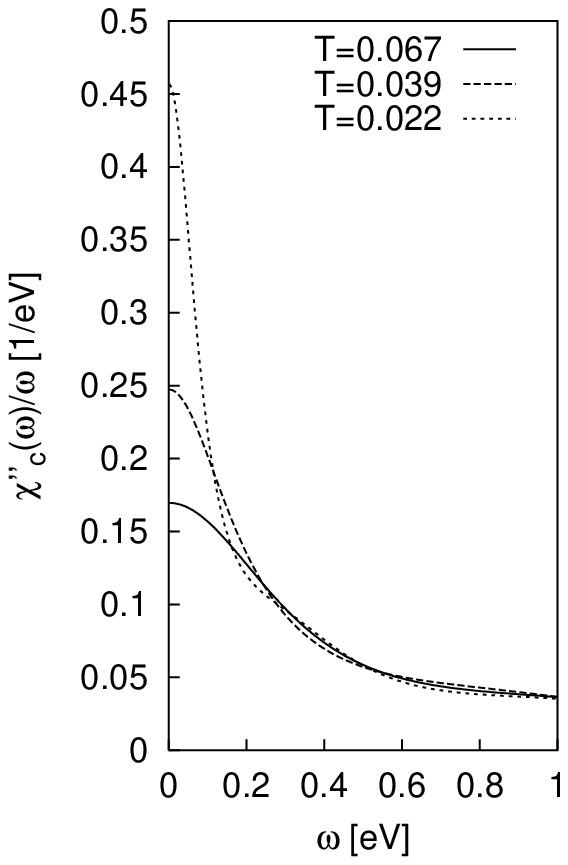}
\caption{The density of states (left) near the Fermi level and the imaginary part of the 
local charge susceptibility over the frequency  (right) at weak doping 
($\delta=0.05$) for different temperatures.  When the 
pseudogap emerges in the density of states, a 
peak develops at zero frequency in the charge susceptibility.}
\label{fig:sus}
\end{figure}

These results are summarized in the temperature-doping ($T$-$\delta$) phase 
diagram shown in Fig.~\ref{fig:pd}. At low doping $\delta$ the system is 
an antiferromagnetic insulator below the Ne\'{e}l-temperature $T_{\rm N}$. 
At finite doping $\delta\leq 0.3$ we find an instability at the critical 
temperature $T_c$ to a superconducting state with a $d$-wave order parameter. 
In the normal state low-energy spin excitations become suppressed below the 
crossover temperature $T^\star$. Simultaneously the electronic excitation 
spectrum represented by the density of states displays a pseudogap, i.e. 
a partial suppression of low-energy spectral weight (see left panel of 
Fig.~\ref{fig:sus}).

In this Letter, we investigate the nature of this transition from the 
normal to the superconducting state and in particular study whether pairing 
in the Hubbard model is driven by the existence of an attractive pairing 
potential as in the BCS theory of superconductivity, or a lowering of the 
kinetic energy. To this end we simulate the superconducting and 
corresponding normal state solutions of the Hubbard model down to 
temperatures $T\approx 0.5T_c$ and compare their respective kinetic 
and potential energies. To obtain the normal state solution we 
suppress superconductivity by not allowing for any symmetry-breaking 
in our representation. 

\begin{figure}
\includegraphics*[width=100mm]{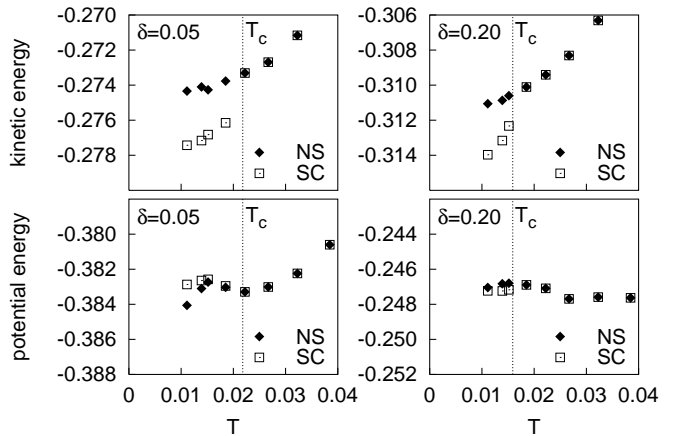}
\caption{Kinetic (top) and potential (bottom) energies of the normal (NS) and 
superconducting state  (SC) as a function of temperature for low doping 
($\delta=0.05$, left) and high doping ($\delta=0.20$, right). The vertical 
dotted lines represent the value of $T_c$. Pairing is mediated by a reduction 
of the kinetic energy.}
\label{fig:E}
\end{figure}

In Fig.~\ref{fig:E} we present the kinetic (top) and potential (bottom) 
energies as a function of temperature at low doping ($\delta=0.05$) on 
the left panel and high doping ($\delta=0.20$) on the right panel. 
The corresponding values of the critical temperatures $T_c$ are indicated 
by the vertical dotted lines. As expected, below $T_c$ the energies of the 
normal and superconducting state start to differ. For both doping levels, 
the kinetic energy of the superconducting state is lower than the kinetic 
energy of the corresponding normal state solution.  This contradicts the 
behavior expected from BCS theory where the kinetic energy of the 
superconducting state is always slightly increased compared to the normal 
state.  In addition, the potential energies of the normal and superconducting states are almost identical, indicating that pairing is not driven by the potential energy.  
The magnitude of the kinetic energy lowering at low doping, measured relative to the 
transition temperature, is roughly 
$\frac{\Delta E_{kin}}{k_{\rm B} T_c} \approx 0.15$, 
in good agreement with the experimental estimate of 
$\frac{\Delta E_{kin}}{k_{\rm B} T_c} \approx \frac{1{\rm{meV}}}{k_{\rm B} 66{\rm K}} =0.15$. 
At $\delta=0.20$, the lowering of the kinetic energy is slightly less compared with $\delta=0.05$.   Thus we conclude that superconductivity in the Hubbard model is driven by a lowering of the kinetic energy with a magnitude that decreases as doping increases.

What could be the underlying microscopic mechanism  for the observed 
kinetic energy driven pairing in HTSC and our simulation? 
Due to the vicinity of the superconducting phase to 
antiferromagnetic ordering, it is widely believed that short-ranged 
antiferromagnetic spin correlations are responsible for pairing in the 
cuprates.   This is the essential idea behind two pairing models which 
predict the experimentally observed lowering in kinetic energy. The first one 
relies on the existence of quasiparticles and is partially based on 
studies \cite{hirsch:hole,bonca:hole,dagotto:hole,barnes:hole} of the motion of  holes in an 
antiferromagnetic background which date back to the early work of Brinkman 
and Rice \cite{brinkman:hole}. The motion of a single hole is inhibited because 
it creates a string of broken antiferromagnetic bonds.  Based on this 
picture, it is argued that two holes can decrease their kinetic energy by 
traveling together, in a coherent motion, i.e.\ by forming Cooper pairs. 
Hirsch's discussion of kinetic energy driven superconductivity 
\cite{hirsch:science} is consistent with this picture.  The second 
idea, due to Anderson, involves spin-charge separation within a resonating 
valence bond (RVB) picture \cite{anderson:rvb}. Due to strong antiferromagnetic correlations, 
spins pair into short-ranged singlets at a temperature $T^*$ much 
higher than the superconducting transition temperature $T_c$. This leads to a pseudogap in the electronic excitation spectrum and consequently to an increase in kinetic energy. Contrary 
to the quasiparticle picture, the elementary excitations of this state 
are spin $1/2$ charge neutral fermions called spinons, and spin 0 bosons 
called holons.  At $T_c$ the holons become coherent and recombine with 
the spinons, forming electrons which pair and render the system 
superconducting. Frustrated kinetic energy is then recovered \cite{baskaran:kineg}.

The first picture relies on the existence of quasiparticles, which in the 
Fermi-liquid concept correspond one to one to with those of a Fermi gas and 
thus have charge and spin.  Anderson's RVB scenario on the other hand is 
based on the concept of spin-charge separation and predicts quasi-free charge 
excitations, the holons.
To distinguish between these two models we investigate the low-energy 
quasiparticle and charge excitations in the Hubbard model by calculating 
the single-particle density of states and the dynamic charge susceptibility,
respectively.  Our result for the density of states in the weakly doped 
system ($\delta=0.05$) for different temperatures above the critical 
temperature $T_c$ is presented in the left panel of Fig.~\ref{fig:sus}. 
As the temperature decreases below the crossover temperature $T^\star$, 
a pseudogap develops in the density of states near the Fermi energy 
($\omega=0$). This partial suppression of low-energy spectral weight clearly 
indicates that no quasiparticles are present in the normal state close 
to the superconducting transition. In the right panel of Fig.~\ref{fig:sus} 
we show the imaginary part of the local dynamic charge-susceptibility 
$\chi^{\prime\prime}_c$  divided by the frequency for different 
temperatures. The low frequency behavior of this quantity provides 
insight in the low energy charge excitations. As the temperature 
decreases, this quantity develops a strong peak at zero frequency, 
indicating the emergence of coherent charge excitations. 

Since the density of states represents quasiparticle excitations which
have both charge and spin, it follows from the simultaneous emergence
of a pseudogap in the density of states and the development of
coherent charge excitations that the low energy spin excitations must
be suppressed.  And indeed, our results for the spin-susceptibility at
the antiferromagnetic wave-vector $(\pi,\pi)$ (not shown) display this
suppression of spin-excitations. Thus, at temperatures below the
crossover temperature $T^\star$ spin and charge degrees of freedom
behave qualitatively different, indicating spin and charge
separation. It is interesting to note that a weak shoulder appears in
the charge susceptibility at $\omega=0.4\approx zJ$, where $z$ is the
coordination number. This observation might be interpreated as a remanence of a residual spin-charge coupling.

\begin{figure*}[h!]
\includegraphics*[width=50mm]{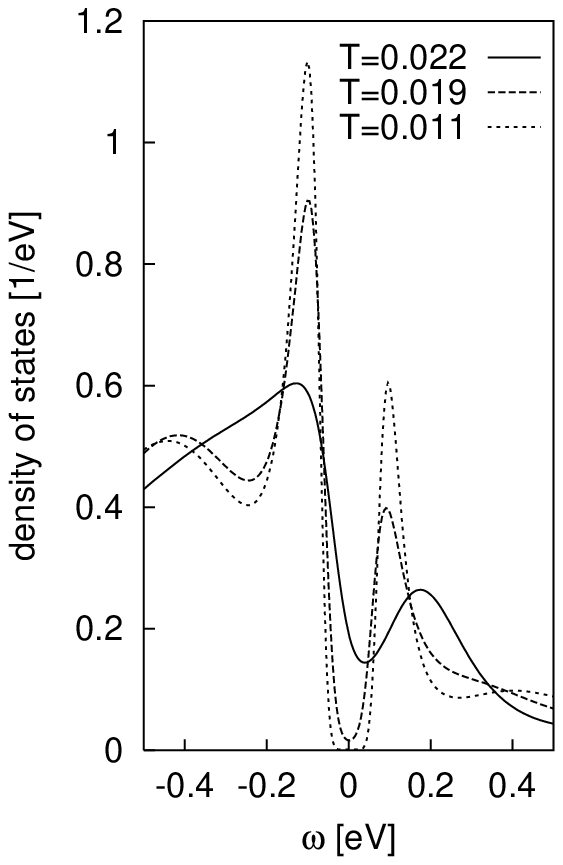}\includegraphics*[width=50mm]{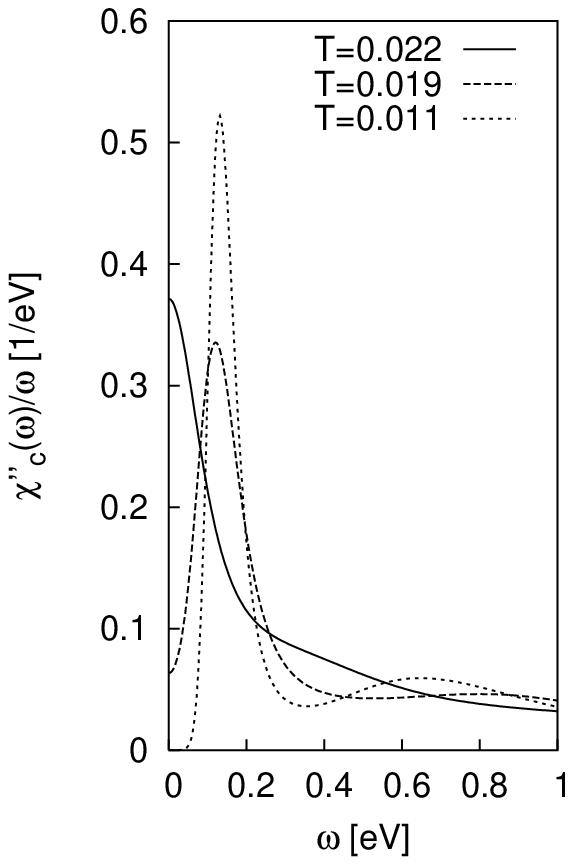}\includegraphics*[width=50mm]{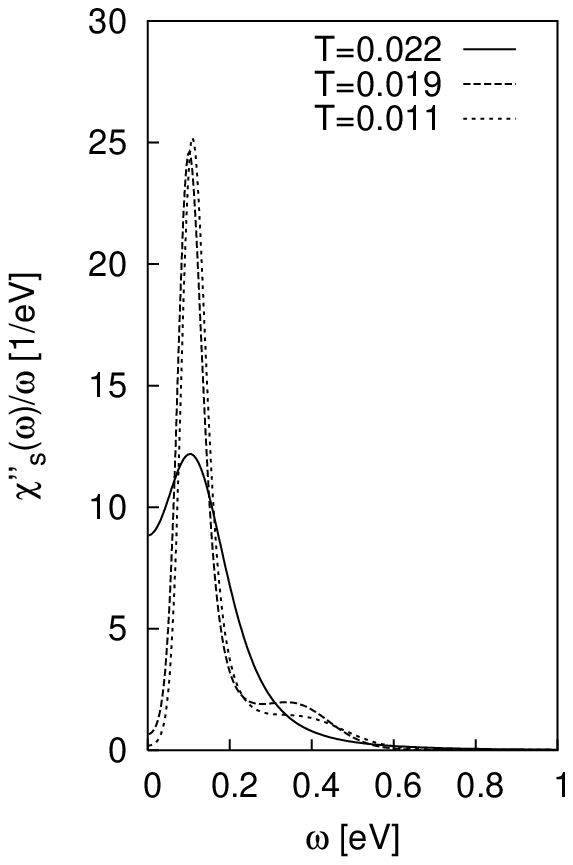}
\caption{The density of states (right), local dynamic charge
susceptibility (center), and the local dynamic spin susceptibility
(right) when $\delta=0.05$, $T_c=0.0218$.  Note that for $T \ll T_c$,
all quantities display a narrow peak delimiting the superconducting
gap, indicating the formation of quasiparticles.  }
\label{fig:sussc}
\end{figure*}

Fig.~\ref{fig:sussc} shows the behavior of the density of states (left
panel), charge- (center panel) and spin-susceptibility (right panel)
at 5\% doping as the temperature decreases below the superconducting
transition temperature $T_c=0.0218$. The density of states and the
spin-susceptibility change smoothly across the superconducting phase
transition. The pseudogap in both quantities changes to a
superconducting gap\footnote{Note that due to the finite resolution in
momentum space, the DCA underestimates low-energy spectral weight in
superconductors where the gap has nodes on the Fermi surface. As a
result we find a fully developed gap at low temperatures instead of a
density of states that vanishes linearly in frequency as expected for
a $d$-wave superconductor.} below $T_c$. However, since the charge
susceptibility is peaked at zero frequency even slightly above $T_c$,
it changes abruptly upon pairing to show the same behavior as the
spin-susceptibility, including the superconducting gap at low
frequencies. Remarkably, well below $T_c$ all quantities display
narrow peaks at $\omega\approx0.1 {\rm eV}$ delimiting the
superconducting gap. This clearly indicates the formation of
quasiparticles below $T_c$.

These results can thus be interpreted within a spin-charge separated
picture as described in Anderson's RVB theory.  The pairing of spins
in singlets below the crossover temperature $T^\star$ results in the
suppression of low-energy spin excitations and consequently in a
pseudogap in the density of states.  The holons, or charge excitations
are free as indicated by the zero-frequency peak in the charge
susceptibility. Well below the transition spin and charge degrees of
freedom recombine, forming electrons which pair.  Frustrated kinetic
energy is recovered as indicated by the reduction of the kinetic
energy as the system goes superconducting.

Using the dynamical cluster approximation we find a kinetic energy driven 
instability in the 2D Hubbard model from an RVB state to a $d$-wave 
superconducting state consistent  with recent optical experiments. 

\paragraph*{Acknowledgements} We acknowledge useful conversations with
G.~Baskaran. This research was supported by the NSF grant DMR-0073308, and 
used resources of the Center for Computational Sciences at Oak Ridge National 
Laboratory, which is supported by the Office of Science of the U.S. Department 
of Energy under Contract No. DE-AC05-00OR22725.

\bibliography{mybib}

\end{document}